\documentclass[useAMS,usenatbib]{mn2e}
\usepackage{graphicx}

\title[Theory of Ring Galaxies ]{Applying the Analytic Theory of Colliding Ring Galaxies }
\author[Curtis Struck]{Curtis Struck$^{1}$\thanks{E-mail: curt@iastate.edu}\\
$^{1}$Dept. of Physics and Astronomy, Iowa State Univ., Ames, IA 50011 USA}

\begin{document}


\pagerange{\pageref{firstpage}--\pageref{lastpage}} \pubyear{2002}

\maketitle

\label{firstpage}

\begin{abstract}
An analytic theory of the waves in colliding ring galaxies was presented some years ago, but the observations where not of sufficient quality then to make quantitative comparisons. Well-resolved observations of a few systems are now available to make such comparisons, and structure imaged in several dozen systems, derived from the recent compilation of Madore, Nelson and Petrillo and the Galaxy Zoo project,  can further constrain the theory. Systems with two rings are especially useful for deriving such constraints. After examining the implications of recent observations of ring sizes and structure,  I extend the analytic theory, investigate limiting cases, and present several levels of approximation. The theory is especially simple in the case of nearly flat rotation curves. I present observational comparisons for a few systems, including: Arp 10, the Cartwheel and AM2136-492. The fit is quite good over a large range of cases. For the Cartwheel there are discrepancies, but the areas of disagreement are suggestive of additional factors, such as multiple collisions. A specific prediction of the theory in the case of nearly flat rotation curves is that the ratio of the outward velocity of successive rings approximately equals the ratio of ring sizes. Ring velocities are also shown to scale simply with local circular velocities in this limit.
\end{abstract}


\section{Introduction: bringing ring galaxy observations, modeling and theory together}

The conceptual theory of colliding ring galaxies as propagating waves, generated by an impulsive disturbance, following a head-on collision was proposed by \citet{bb18}, and supported by numerical models. One of the nice properties of the Lynds and Toomre ring galaxy theory is that it has a perturbation limit. When the companion-to-primary galaxy mass ratio is small, or the relative velocity is large, the impulsive disturbance will be small. In this case, the structure of the target galaxy, including its global gravitational potential and its flattened disc, will not be greatly perturbed. This fact provides a foundation for approximate treatments of the ring waves.

An analytic model for stellar rings based on this approximation, and making use of the theory of caustic wavefronts, was developed by the present author and collaborators in several papers beginning in the late 1980s (e.g., \citet{bb28}, \citet{bb26}). This and other work published by the mid-1990s on colliding ring galaxies was summarized in the review article of \citet[henceforth ASM]{bb3}. Since that time a number of new numerical models have been published (see discussion and references in Sec. 4). These have generally confirmed the early conceptual theory, and provided new extensions. Most of the results of these numerical models have been in accord with the qualitative predictions of the analytic model. However, few detailed comparisons have been made to date.  

The nonlinear dynamics of most galaxy collisions make quantitative analytic modeling impossible. In the cases where it is possible there could be substantial benefits to comparing these predictions to observations and numerical models. Where the analytic models and observations agree numerical models can be checked. Where numerical and analytic models agree they can provide powerful tools for interpreting high resolution observations, and extending the theory. I will provide specific examples of these general statements in Sec 4.  

In the case of colliding ring galaxies there have been several barriers to the quantitative application of the analytic model since its development. The first barrier is that Ð it has not been clear whether the perturbation approximation is applicable to any of the (few) well-studied ring galaxies. This barrier may be best overcome by ignoring it, i.e. testing the application of the analytic model. However, the second barrier is more substantial and practical, except for a very small number of cases, like the well-known Cartwheel ring, sufficient observational data to allow detailed comparisons have not been available. Moreover, the best-studied cases (like the Cartwheel, see Sec. 4.2) are not the most symmetric rings, and often their companions are not small enough to obviously fit the perturbation approximation. However, much more data has been acquired in the last decade. Examples are given in Sections 2 and 4. 

The final barrier to the application of analytic rings theory was also practical. As originally formulated the analytic models had a number of parameters, and some of the qualitative behaviors depended sensitively on those parameters. Moreover, the equations were moderately complex and the formalism of caustic waveforms is not a commonplace in astronomy. In retrospect, the development of the formalism, e.g., as summarized in ASM, may have been overly broad. In the time since a great deal has been learned about the universal scaling properties of all types of galaxies. Observational surveys of large samples of galaxies like the SDSS have allowed statistically meaningful average properties and dispersions around them to be determined. Large scale cosmological simulations of galaxy formation, like the Millennium simulation \citep{bb24}, have greatly improved our understanding of halo formation and buildup. This, in turn, has allowed the testing of universal analytic halo profiles against the numerical models. 

For the purposes of this paper, the fundamental result of most interest from among these advances is the universality of flat rotation curves in galaxy discs. This allows us to considerably restrict the range of halo profiles for ring galaxies and their companions. In essence we can make a second perturbation approximation -  that the flat rotation curve potential is the standard, and others are generally a small perturbation from it. We will see in Sec. 3 that this approximation allows us to considerably simplify the analytic ring wave equations, essentially eliminate some parameters, and derive some very specific quantitative predictions that can be compared to the new generation of observations.  The generally favorable result of this comparison implies that we have a very complete understanding of the simplest kind of galaxy collision, and holds out the hope that analytic models can be used to extend this understanding to less symmetric cases.

\section[]{Observational ring phenomenology}

When the ASM review was written only one symmetric ring galaxy had been studied in detail with high resolution, multi-waveband observations $-$ the Cartwheel. Even then, there were indications that the Cartwheel was not the most representative of the class. This will be discussed further in Sec. 4. Meaningful comparisons between theory and observation require extensive data from at least a few representative systems, and moderately resolved optical observations of a much larger sample to determine average properties. Acquiring such data is especially difficult for ring galaxies because they are a rare subset of a rare class of galaxy (galaxies involved in major collisions). Only a modest number have been discovered in surveys, and many of these are quite distant by the standards of low redshift studies. 

As we will see in Sec. 4, detailed studies have been carried out on several prominent systems, including HI density and kinematic mapping. New imaging collections have also been published. The first such collections were derived from the \citet{bb4} and \citet{bb5} atlases of peculiar galaxies. \citet{bb13} assembled an extensive list of ring galaxies from the latter atlas. However, many of these objects were not colliding ring galaxies, and images were not readily available for most of the list objects. Recently, \citet[henceforth MNP]{bb19} have provided a refined list and images of about 100 probable colliding rings, with the selection based on collision morphologies and the presence of a possible collision partner. This very helpful resource motivated the present study. A collection of multi-colour observations of more than a dozen rings was also recently published by \citet[also see the study of \citealt{bb2}]{bb22}. More multicolour images have been produced by the Sloan Digital Sky Survey (SDSS), and classified by the Galaxy Zoo project on the internet.

As will be discussed in the next section, among the key predictions of the analytic rings theory are the relative sizes of successive rings and stellar ring widths. The ring widths are still difficult to analyze since images of most systems are only able to detect the young stars formed from the gas component of the ring. Historically, successive ring sizes have been hard to study because very few systems with multiple rings have been discovered. The MNP collection and Galaxy Zoo objects provide some new, or at least relatively unknown, examples, though the number remains very small. 

However, the analytic theory shows that second and third rings will generally be broader than the first. This suggests that it will take longer for them to clearly separate from an inner bulge or bar component. If the disc gas is dispersed by strong star formation in the first ring wave, then star formation may be weaker in subsequent waves, which would also make them harder to detect. On the other hand, the surface brightness in the inner regions of many ring galaxies is apparently rather sharply truncated at an outer edge, suggesting that we may be seeing the outer edge of a second (or later) ring rather than a typical declining bulge profile. In some cases, a partial (inner) ring arc is seen. The Cartwheel, without a large obscuring bulge, appears to a good example of these generalizations. In any case, the outer edge of the inner light distribution provides an upper limit to the size of an inner ring, and even limits on inner ring size relative to the outer ring size can be usefully compared to the theory.

In the next subsection I will present a subset of the MNP rings, with a few additions, for which such ring size comparisons can be made. The criteria for inclusion in this subset are the following. 

1. The ring is reasonably symmetric, elliptical, and the majority of it is visible. These symmetry conditions are designed to eliminate very warped rings or ring-like spirals, which are the result of additional interaction variables beyond those considered by the simple analytic theory. The visibility criterion is designed to eliminate systems where most of the ring image is so faint it is hard to be certain that it is in fact a symmetric ring produced by a direct, symmetric collision. A large fraction of the MNP rings suffer from such symmetry/visibility limitations. This is especially true when only a single sky survey image is available for the system. 

2. A related, but more specific condition is that it must be possible to fit an ellipse to the outer ring. In principle, this criterion eliminates Ôrings,Õ whose oval shape is not a circle or projected circle. In practice, such cases are rare, though in a few cases I have allowed a very liberal interpretation of the ellipse fit. The most egregious example is Arp 143, the ÔtriangularÕ ring. 

3. Another related condition is that the ring thickness must not vary greatly with azimuth. Generally, this circumstance is the result of an off-centre collision, which will introduce torques not considered in the simple theory (see ASM). I have also relaxed the application of this restriction in some cases, but a good example of a beautiful colliding ring that was eliminated with this condition is the ``Sacred Mushroom'' AM 1724-622 \citep{bb30}.

4. A stringent requirement is that in addition to the outer ring, an inner luminosity component must be present. This eliminates many empty rings, like the western component of Arp 147, a beautiful image of which was recently produced by the Hubble Heritage Project. Most empty rings probably come from late-type progenitors, without a substantial bulge or bar, and have a companion that is substantial enough to destroy a nuclear star cluster and central gas disc (see the Hubble Heritage images of Arp 147 and Arp 148). This condition also eliminates many systems that are too young and small to have fully detached from their central regions. The main reason for imposing this restriction is that it is a necessary pre-condition for having an inner ring in addition to an outer ring. Of course, in most cases the inner component will be a bulge, a bar, or a ring plus bulge or bar, rather than simply a nascent inner ring. For the reasons given above the different possibilities generally cannot be distinguished without much higher resolution observations than presently available. 

5.  A final symmetry requirement is that the central light distribution should not be displaced too far from the ring centre. Specifically, I generally require that the central light core be closer to the ring centre than to the ring itself. This requirement eliminates some very interesting galaxies, like NGC 985. However, the rule has been stretched to include the Lindsey-Shapley ring (AM0644-741) and VII Zw 466.

\subsection{The MNP subsample}

Application of these conditions reduces the MNP to a couple of dozen objects, as listed, with the addition of a few non-MNP objects, in Table 1. This is our primary sample of possible multiple ring galaxies. 

\begin{table*}
 \centering
 \begin{minipage}{140mm}
  \caption{Ring diameters.}
  \begin{tabular}{@{}lccccl@{}}
  \hline
  {Ring System  \footnote{Systems listed in the order of MNP, i.e., R.A. order for Arp-Madore systems followed by northern systems.}} &  Outer ring & Outer ring & Ring size & 
  {${r_2}/{r_3}$ \footnote{These are based on the authorÕs measurements, from various published images. }}
  & Comment\\
 & diameter & diameter & ratio & &\\
 & {(arcsec) 
\footnote{All values from MNP, except for II Hz 4, which is from \citet{bb20}. All values estimated from longest axis.}} 
 & {(kpc) 
 \footnote{Values derived from previous column and (Galactocentric, GSR) distance given in NED.}} 
 & ${r_1}/{r_2}$ & &\\
 
 \hline
 1. Cartwheel & 58 & 35 & 4.2 & 1.8 &\\
   AM0035-335 &&&&&\\
 2. AM0425-421 & 55 & 16 & $>$ 2.1 & - &\\
 3. AM0643-462 & 60 & 47 & $>$ 3.2 &-& Like AM0644-741\\
 4. AM0644-741 & 97 & 41 & $>$ 3.2 &-& Lindsay-Shapley ring\\
 5. AM1133-245 & 67 & 51 & $>$ 2.6 &-&\\
 6. AM1135-284 & 24 & 8.6 & $>$ 2.3 &-&\\
 7. AM 1323-222 & 88 & 26 & 2.0 & $>$ 1.8 & {Rings unclear 
 \footnote{On the available image the rings are quite indistinct. They could be rings 2, 3, 4 with small spacings, as assumed here. Alternately, the outer ring could be ring 1, and the inner rings could be the inner and outer edges of a wide ring 2.}}\\
 8. AM1354-250 & 53 & 21 & $>$ 2.9 &-&\\
 9. AM1358-221	& 53	& 38 & $>$ 3.8 &-& The western ring\\
 10. AM1413-243 & 35 & 32 & $>$ 2.8 &-&\\
 11. AM1434-783 & 40 & 12 & 3.5 &-& Spokes?\\
 12. AM2100-725 & 27 & - & $>$ 2.2 & - &\\
 13. AM2132-535 & 32 & - & $>$ 3.3 & - &\\
 14. AM2136-492 & 72 & 76 & 1.8 & $>$ 2.1 & Bulge or third ring?\\
 15. AM2200-715 & 32 & - & $>$ 2.5 &-& Bar or second ring?\\
 16. AM2230-481 & 52 & 36 & $>$ 3.2 & - &\\
 17. AM2238-541 & 32 & - & $>$ 1.8 & - & Colliding or just barred?\\
 18. Arp 318 & 40 & 11 & $>$ 2.3 &-&\\
 19. Arp 10 &  44 & 27 & $>$ 3.3 &-&\\
 20. Arp 147 & 17 & 11 & 3.2 &-& East galaxy. Second\\
 &&&&& ring uncertain.\\
 21. ESO 200 - & 29 & 33 & $>$ 2.1 &-&\\
 IG009 &&&&&\\
 22. Arp 143 & 87 & 23 & $>$ 3.3 &-& Messy ring.\\
 23. NGC 2793 & 46 & 5.1 & $>$ 4.0 &-& Bar, not second ring.\\
 24. IC 0614 & 34 & 23 & $>$ 2.1 &-&\\
 25. VII Zw 466 & 21 & 20 & 3.0 &-& Very off-centre second ring.\\
 26. II Hz 4 & 28 & 24 & 3.0 &-& Emerging second ring?\\

\hline
\end{tabular}
\end{minipage}
\end{table*}

Two or three ellipses were fitted (manually, by eye) to each object in this sample. In most cases, survey images from MNP were used. These were originally derived from IIIa-J photographic plates from the UK Schmidt southern sky survey. When superior optical images were available, e.g., via the NASA Extragalactic Database, they were used. Multi-colour SDSS or Hubble Heritage images of several systems were found there. In each case, the first ellipse was fitted to the middle of the outer ring. I will be referring to the ``middle'' or central ridge line of finite width rings many times in this paper. To simply, I will henceforth call this the RCC for ring central circle. A second, smaller ellipse was fit either to the visible inner ring, or to the apparent outer edge of the inner light distribution. In the former case, a third ellipse was fit to the apparent outer edge of the inner light distribution. An example of the fitted ellipses is shown in Figure 1. The ring sizes given in Table 1 are generally very close to those of MNP, I measured them all for internal consistency with my measurements of sample objects that were not included in MNP.

\begin{figure}
\includegraphics[width=88mm]{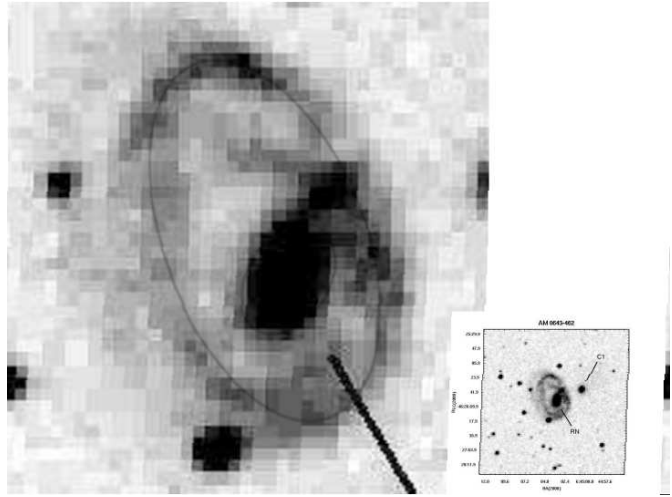}
 \caption{An example of the ellipse fitting procedure. The inset shows the image from Fig. 5 of MNP; the expanded image shows the by eye ellipse fitting to the central ridge of the outer ring, and the outer edge of inner luminous region. The semi-major axis of the latter is taken as an upper limit to the inner ring radius.}
\label{fig1}
\end{figure}

Generally, the innermost ellipse will bound the outer edge of the innermost ring. When it actually coincides with the outer edge of the innermost ring, it provides an overestimate of the size of that ring. In most cases the innermost ring is either buried within a bulge or a bar, has not formed, or will not form. The latter case is most likely if the innermost ring would be a third or higher order ring, where phase mixing prevents the development of any distinguishable ring.  In all of these three cases the ellipse is substantially bigger than the innermost ring, and the ratio of next larger to the inner ellipse diameter will be a lower limit to the corresponding ring size ratio. 

Nonetheless, it is possible that the innermost ring is larger than the central light distribution, but is so faint that it is not detected on the available image. In this case we will underestimate the inner ring size, and the ring size ratio will be too large. Since the ring waves will generally incorporate most of the stars in their radial domain, this circumstance seems very unlikely unless the stellar surface density falls unusually rapidly within the disc. However, it could occur if, as in low surface brightness galaxies, the stars are confined to a small central region, and the disc consists mostly of gas. The outer ring could induce star formation, making it visible, while the inner ring could have propagated past the stellar core, and been too weak to induce star formation. Again, this would seem to be a rare happenstance.

Another important assumption was adopted in estimating the ring size ratios of Table 1. It was assumed that the longest axis of the ellipse represented the size of the ring, as projected onto the plane of the sky. The same assumption was made for inner ring(s). This assumption is based on the notion that we have selected symmetric rings whose true form is circular. Even in this restricted sample there is evidence that this is not always the case, and that the shape of some rings is due to more than just projection of a circle. This evidence includes nuclei offset from the ring centre, and a lack of alignment between the two ellipses. Extremes of the former were not included in the sample. In most systems the ellipses are fairly well aligned, but a significant fraction have large misalignments. 

In most of these last cases we do not actually resolve an inner ring, so the inner ellipse may only be outlining a bar component. An exception is the Cartwheel, which has resolved and misaligned rings.

Figure 2 shows the computed ring ratios versus object number in Table 1. The most obvious result apparent in Figure 1 is that there are no ratios less than about 1.8 and none above the 4.2 of the Cartwheel. Granted, most of the points are lower limits, so that there could well be rings with higher ratios. The points representing lower limits are scattered throughout the range between these two limits.  Interestingly, all of the points seem to concentrate around values slightly greater than 3.0 and about 2.0, and this is especially true of the points representing apparently resolved multiple rings. However, beyond the fact that the points cluster between values of 1.5 and 4.5, these results are not very statistically significant. This is especially true given all the caveats above about the systems with unresolved inner rings. Yet when compared to the analytic theory below, they are very suggestive, and certainly provide motivation for obtaining well-resolved observations of more ring galaxies. 

As an aside we note that in the spiral sample recently studied by \citet{bb21} the diameter ratios of the outer Lindblad resonance to the inner Lindblad resonance similarly range over factors of about 2-3. Since both types of wave are associated with epicyclic motions this is not entirely surprising. We also recall that \citet{bb5a} studied the statistical properties of double ringed galaxies from the large sample of  \citet{bb11a}. They found that in barred galaxies the average value of the ring diameter ratio was about 2.2, while this value ranged more widely (from about 1.7 to 11.7) in non-barred galaxies. In most cases these rings are probably located at Lindblad resonances in isolated galaxies. These examples emphasize the nature of rings cannot be determined from ring ratios alone. Kinematic evidence or evidence of interactions are also needed.

\begin{figure}
\includegraphics[width=90mm]{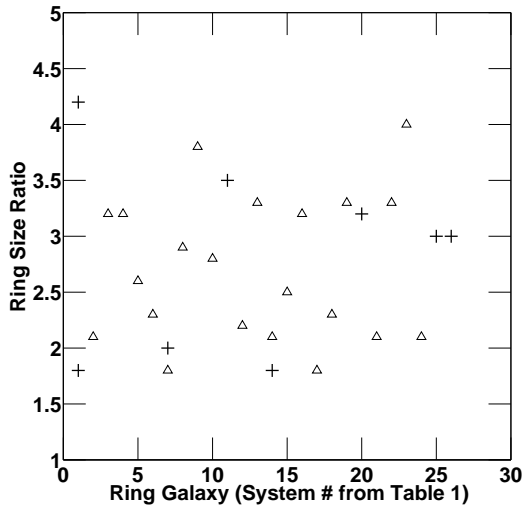}
 \caption{Ring size ratio versus object number from Table 1. Plus signs mark measured values, upward pointing triangles give lower limits. }
\label{fig2}
\end{figure}

\subsection{The Galaxy Zoo sample}

The Arp and Arp-Madore catalogs illustrate the value of careful searches of survey images to discover examples of rare galaxy types, like the colliding rings. The Sloan Digital Sky Survey, with images of many millions of galaxies presents a great opportunity for such work, but one beyond the capabilities of any small group of investigators. Fortunately, the Galaxy Zoo project has mustered a much larger group of amateur investigators to do the sorting. There are two levels of sorting in the project. The first is the general classification carried out by any interested participant. The second is a more informal sort by participants of specialized internet forums within the project. One of these focusses on colliding ring galaxies. 

From the many objects accumulated in that forum by October 2009 I've selected a dozen for a second sample, see Table 2. These objects (including one found by the author in the course of general classification) generally satisfy the 5 criteria applied to the MNP sample above. Of course, the MNP sample is already a select subset of the \citet{bb13} sample designed to increase the likelihood of them being colliding rings. Thus, I added a couple of selection criteria like those used in the MNP sample. Specifically,``theta-rings,'' with a bar-like structure extending across the ring were excluded, and each object had to have a plausible companion within a few diameters. Here plausible means of appropriate size so that is was not likely a foreground or background object, and was substantial enough to be a possible cause of the ring wave. 

Table 2 lists the SDSS designations of these objects and alternate names. The latter show that most of these objects could already be found in earlier catalogs. However, in many cases images of the quality of the SDSS are needed to recognize them as ring galaxies. Table 2 also gives ring diameters and ratios determined as described in the previous subsection. The ring size ratios and limits for this sample are shown in Figure 3. The range of values there is very similar to that of Figure 2, given the limited numbers in both samples. Yet while this second sample is smaller than the first, it has a larger fraction of apparently resolved inner rings. Thus, there are more crosses in Fig. 3 than Fig. 2, and these confirm that the typical size ratio of the first to second ring is about 3. 

\begin{table*}
 \centering
 \begin{minipage}{140mm}
  \caption{Galaxy Zoo sample ring diameters.}
  \begin{tabular}{@{}lccccl@{}}
  \hline
  Ring System &  Outer ring & Outer ring & Ring size & 
  {${r_2}/{r_3}$ \footnote{These are based on the authorÕs measurements, from SDSS images. }}
  & Comment\\
 & diameter & diameter & ratio & &\\
 & {(arcsec) 
\footnote{ All values estimated from longest axis.}} 
 & {(kpc) 
 \footnote{Values derived from previous column and (Galactocentric, GSR) distance given in NED.}} 
 & ${r_1}/{r_2}$ & &\\
 
 \hline
1. J090225.39 & 33 & 53 & $>$ 2.4 &-&\\
  +553633.2 &&&&&\\
2. J092603.26 & 31 & 18 & 2.9 & $>$ 1.7 &\\
  +124403.7&&&&&\\
  (UGC 05025)&&&&&\\
3. J105007.28 & 52 & 25 & $>$ 3.3 &-&\\
  +362030.5&&&&&\\
  (UGC 05936)&&&&&\\
4. J110003.87 & 43 & 25 & 3.1 & $>$ 2.7&\\
  +172527.6&&&&&\\
  (PGC 033141)&&&&&\\
5. J125255.13 & 67 & 51 & $>$ 2.6 &-& Two double rings \\
  +320451.2&&&&& {in system.\footnote{In the primary, the two rings, and the outer edge of the presumed third are all aligned. The outer ring of the companion is very faint.}}\\
  (KUG 1250+323)&&&&& See next row.\\
6. J125302.93 & 34 & 16 & 3.2 & $>$ 2.8 &Diameter in kpc \\
  +320625.2&&&&&assumes same\\
  (KUG 1250+323)&&&&&distance as 5.\\
7. J133547.36 & 18 & 22 & 2.4 & $>$ 2.4&Broad, faint outer\\
  +455037.5&&&&&ring.\\
8. J145556.45 & 23 & 10 & 4.6 & $>$ 3.3 &Wide outer ring.\\
  +115229.4&&&&&\\
  (KPG 445B)&&&&&\\
9. J153227.65	& 28	& 23 & 3.0 & $>$ 2.2&\\
  +414842.3&&&&&\\
  (CGCG 222-022)&&&&&\\
10. J160153.01 & 39 & 33 & 2.5 & $>$ 2.5 &\\
  +452107.0&&&&&\\
  (PGC 056751)&&&&&\\
11. J172430.82 & 42 & 26 & $>$ 2.9 &-&\\
  +565434.9&&&&&\\
  (CGCG 277-042)&&&&&\\
12. J230658.93 & 78 & 33 & 2.8 & $>$ 2.3& NGC 7489 is\\
  +225611.3&&&&&nearby, with \\
(IC 5285)&&&&&similar redshift.\\
\hline
\end{tabular}
\end{minipage}
\end{table*}

\begin{figure}
\includegraphics[width=90mm]{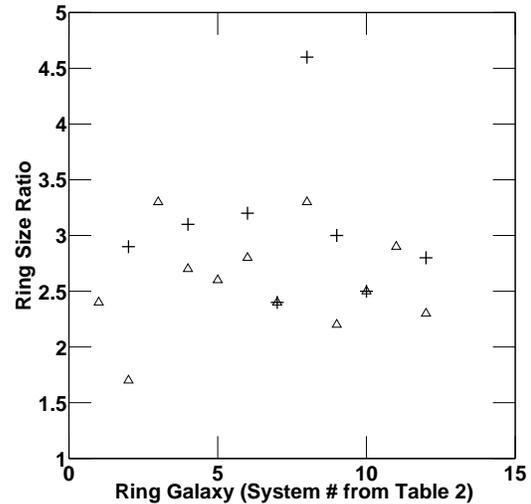}
 \caption{Ring size ratio versus object number from Table 2. As in Fig. 2 plus signs mark measured values, upward pointing triangles give lower limits. }
\label{fig3}
\end{figure}

\section{Simple analytic theory and applications}

In order to understand the suggestive results of the preceding section, and provide more specific predictions to guide future observations, we reexamine the analytic theory in this section. This discussion closely parallels that of Sec. 4.2.1 of ASM, with modifications and extensions. The analytic theory is based on several fundamental approximations: 1. the collision is perfectly symmetric (i.e., the orbit of the companion is along the rotation axis of the target disc), 2. the Impulse Approximation describes the disturbance, and 3. the epicyclic approximation gives the post-collision motion of stars in the ring galaxy disc. An ancillary approximation is that the target disc is not significantly disturbed in the direction perpendicular to its initial plane. Additionally, we will not consider gas dynamical effects. Alternatives to the impulse approximation were considered in ASM, and the analytic precessing elliptical orbits considered in \citet{bb25} could be used in place of epicyclic orbits. However, in neither case would gains in accuracy offset the added complications to the formalism. 

In this section I will partially re-interpret the ASM discussion by focusing on the limiting case of flat rotation curve (FRC) galaxies, both the ring galaxy and its companion. The analytic formalism is particularly simple in this limit, which in addition to its physical significance to galaxies, means that it provides a good baseline state for comparing other nearby states (modestly rising or falling rotation curves). We begin by considering the amplitude of the impulsive disturbance that drives the waves.

\subsection{Disturbance amplitudes in flat rotation curve galaxies}

Before the collision we assume that all orbits in the target disc are circular. According to the Impulse Approximation, immediately after the collision the stellar orbits are unchanged except for the addition of a radial velocity component resulting from the inward acceleration due to the companion's gravity. After the collision, this component is given by,

\begin{equation}
{v_r}(t) = {\Delta}{v_r}\ \cos{({\kappa}(q)t)},
\end{equation}

\noindent
where ${\Delta}v_r$ is the impulsive disturbance, $q$ is the initial unperturbed radius of the given star, and ${\kappa}(q)$ is the radial epicyclic frequency at radius $q$. The velocity impulse is approximately given by,

\begin{equation}
{\Delta}{v_r} = a{\Delta}t 
= {\frac{-G{M_c}(q)}{q^2}} {\Delta}t,
\end{equation}

\noindent
where $a$ is the average acceleration on a target disc star, ${M_c}(q)$ is the companion mass interior to the radius $q$, and $\Delta t$ is the time interval that the companion spends within a distance $q/2$ of the target centre, exerting this pull on the disc star. Then, $\Delta t \approx q/v_{rel}$ where $v_{rel}$ is the relative velocity of the galaxies at impact, and is assumed constant over the relatively brief time interval ${\Delta}t$. If the companion galaxy is an FRC galaxy (like the target), with a circular velocity of $v_{cir2}$, then we have,

\begin{equation}
{\Delta}{v_r} =  {\frac{-v^2_{cir2}}{q}} \Delta t
= {\frac{-v^2_{cir2}}{v_{rel}}} = constant.
\end{equation}

\noindent
Because for a flat rotation curve  $\kappa \sim 1/q$, the relative amplitude $A = (-\Delta v_r)/(q \kappa)$ is also constant. Both conceptually and for formal manipulations this is a great simplification.

\subsection{Finding ring radii}

For the moment we will set aside the result of the last subsection that in the FRC case (both galaxies) the disturbance amplitude is independent of radius in order to present the formalism in a more general context. Following ASM (Eq. 4.1) we can approximate the rotation curve of the target galaxy as a power-law,

\begin{equation}
v_{cir} = v_{\gamma} \left( \frac{r}{\gamma} \right)^{1/n},
\end{equation}

\noindent
where $\gamma$ is a reference radius, and $v_{\gamma}$ is the azimuthal velocity at that radius. I will assume that the same form applies to the rotation curve of the companion, but with a different power-law index ($n'$) and a different scale velocity ($v_{{\gamma}c}$).

In the theory of ASM (and \citealt{bb28}) nonlinear ring waves are bounded by sharp inner and outer edges where orbiting stars pile up before reversing their radial motion. The ring itself is a region of enhanced stellar density where stellar orbits cross, which they do not do in the rarefied zones outside the rings. Formally, the ring edges are caustics where the density goes to infinity. These caustics are defined as regions where the first-order derivative of radius with respect to initial radius (${\partial}r/{\partial}q$) goes to zero. (I.e., the radial compression goes to infinity.) Generally, the caustics are born at a finite radius, where the two edges meet in a cusp, and ${\partial}r/{\partial}q = {\partial}^2 r/{\partial}^2 q = 0$. At times after the cusp appears, these two equalities are satisfied at different radii. Specifically, the second derivative condition defines the centre of the ring, while the first derivative condition defines the inner and outer limits. For nearly FRC cases the ring width is not large, however.

For estimating the relative sizes of different rings we are most interested in the location of the RCCs (ring central circles) of finite width rings , and thus, in the second derivative condition. It can be written (see ASM), 

\begin{eqnarray}
\nonumber
\left( \frac{2}{n'} \right)  \kappa t\  \cos{(\kappa t)} -
\left( \frac{1}{n} - 1 \right) (\kappa t)^{2}\ \sin{(\kappa t)} \\
\nonumber
= \frac{\left( {\frac{1}{n}} - {\frac{2}{n'}} \right)} {\left( \frac{1}{n} - 1 \right)}
\left( \frac{q \kappa}{{|\Delta} v_r|} \right) 
= \frac{\left( {\frac{1}{n}} - {\frac{2}{n'}} \right)} {\left( \frac{1}{n} - 1 \right)}
\frac{1}{A_{\gamma}}   \left( \frac{q}{\gamma} \right)^{{\frac{1}{n}} - {\frac{2}{n'}}},\\
\nonumber\\
\mbox{with}\ 
{A_{\gamma}} = \left( \frac{|{\Delta} v_r|}{q \kappa} \right)_{q=\gamma},
\end{eqnarray}

\noindent
where again $n'$ is the rotation curve index of the companion, corresponding to $n$ in Eq. (4) for the primary. When the rotation curve is not flat the impulsive perturbation $\Delta v_r$ in the target disc varies with radius, and is a power-law for a power-law rotation curve. Specifically, Eq. (2) implies that $\Delta v_r \sim q^{2/n'}$, which accounts for the final factor in Eq. (5). To simplify the last exponent in Eq. (5) it is convenient to define,

\begin{equation}
\frac{1}{n''} = \frac{1}{n} - \frac{2}{n'}.
\end{equation}

\noindent
Equation (5) appears complex. However, we will see that in physically relevant cases it is quite manageable.

\subsection{Ring sizes in the FRC case}

When the rotation curves of the two galaxies are flat, $n$,$n'$ and $n''$ go to infinity. In this limit, Eq. (5) reduces to the very simple form

\begin{equation}
(\kappa t)^2 \sin{(\kappa t)} = 0.
\end{equation}

\noindent
Then, the RCCs of successive rings are found at epicyclic phases of  $\kappa t = \pi, 3\pi, 5\pi...$ At these phases the radii of ring RCCs are the same as their unperturbed radii, that is, $r =q -Aq \sin{(\kappa t)} = q$ (where $A$ is given by the last of Eqs. (5), but here for any $q$ value). In this case $\kappa = \sqrt{2} v_{cir}/q$, so the radii of successive rings are given by

\begin{equation}
q_{ri} = \frac{\sqrt{2}v_{cir}t}{(2i-1)\pi},   \ \ 
\mbox{for}\  
i = 1, 2, 3...
\end{equation}

\noindent
and the relative sizes of successive rings are given by,

\begin{equation}
\frac{q_{r1}}{q_{r2}} = 3,
\frac{q_{r2}}{q_{r3}} = \frac{5}{3},...
\frac{q_{ri}}{q_{r(i+1)}} = \frac{2i+1}{2i-1}.
\end{equation}

\noindent
The remarkable simplicity of this case, and this result in particular, was not appreciated previously. In non-FRC cases the term on the right-hand-side of Eq. (5) does not disappear. The extra dependence on q in that term means that the ratios of ring sizes will vary with time or with the ring size. Moreover, in such cases the RCC phase, $\kappa t$, of a ring will not be a multiple of $\pi$, and the $r-q$ relation will also depend on time or additional factors in $q$. There are also dependences on the parameters, $n$, $n'$, and $A$ in these cases.

On the other hand, most galaxy rotation curves are only moderately rising or falling if they are not flat, so all these additional dependences may not have a great effect, and the FRC case may be a good guide. Nonetheless, before attempting to compare to observation, it is worth considering what effect they do have on the ring size ratios.

\subsection{Approximate ring sizes in non-FRC cases}

The discussion of the previous section suggests that in analyzing Eq. (5) we should consider $\kappa t$ as the primary variable, and the other variables as parameters, which when varied over reasonable ranges have a moderate effect on solutions $\kappa t$. In essence, the question to consider is how far can the $\kappa t$ values for RCCs vary from (odd) multiples of $\pi$, since this determines the ring positions and size ratios (i.e., via generalized versions of Eqs. (8) and (9))?

We can begin to answer this question by considering the ranges of the rotation curve exponents $n$ and $n'$. First, note that $n = 1$ is the solid body case. Rings do not form in the solid body case. Values of $n$ between 0 and 1 imply a rotation curve that rises more steeply than solid body, and yield unphysical inward propagating waves (see ASM). Since rotation curves do not generally rise very steeply outside of core regions of galaxies, it seems reasonable to restrict consideration to positive values of $n$ (or $n'$) greater than a few.

Negative values of $n$ and $n'$ correspond to falling rotation curves. The Keplerian case corresponds to $n = -2$. In general, we expect falling rotation curves in galaxies to be much less steep than that, so we can expand our restrictions to absolute values of $n$ (or $n'$) greater than a few. By the definition of $n''$, its value will be similarly constrained, and the $q$-dependence on the right-hand-side of Eq. (5) will always be weak. 

There are limits on the constant terms on the right-hand-side of Eq. (5) as well. The amplitude $A$ must be greater than about 0.1 in order for there to be a significant ring wave, and the constant factors will be of order unity or less, depending on the value of $n'$. 

When $|n|, |n'| > 10$, the second term on the left-hand-side of Eq. (5) must be small because the all other terms in the equality are. The solution for $\kappa t$ must be close to a multiple of $\pi$ as in the FRC case. Only when these exponents take values between 3 and 10 (or their negatives) do we expect any significant deviation from the FRC case, subject to the adopted restrictions. 

This kind of argument can be pursued from another direction, and the conclusion can be generalized.  We begin with a generalization of Eq. (9) for the ratios of ring radii to the non-FRC case,

\begin{equation}
\frac{q_{r1}}{q_{r2}} = 
\left( \frac{\phi_{r2}}{\phi_{r1}} \right)^{1-1/n},
\end{equation}

\noindent
where I adopt the terms $\phi_{r1}$ and $\phi_{r2}$ for the phases $\kappa t$ of the RCC of the first and second ring waves. These phases are odd multiples of $\pi$ in the FRC case, but need not be in general. Also in the FRC case, $\phi_{r2} = 2\pi + \phi_{r1}$. Since the RCC phase of any ring will depend on $q$ in the general case, this relation will also not be adhered to. However, we expect that it will be approximately true since the additional $q$-dependences of Eq. (5) in the non-FRC cases are moderate. Nonetheless, even small differences in its value can have a significant effect on the analytic solution.

Generally, the first-to-second ring ratio will be large when $\phi_{r1}$ is small. It is physically implausible that $\phi_{r1} < \pi/2$, i.e., where the ring centre would be an at epicyclic phase corresponding to the initial radial infall of stars. In this extreme case the ring ratio (with $\phi_{r2} = 2\pi + \phi_{r1}$) would be less than 5.0, depending on the value of $n$. At a more moderate phase of $3\pi/4$ (i.e., in terms of deviations from the canonical phase of $\pi$), the ring ratio would be less than 3.7, not much greater than 3.0.

Taking another point of view we can ask, what does it take to get a ring phase of about $\pi/2 < \phi_{r1} < 3\pi/4$, which is needed to get a large ring ratio? This question leads us back to Eq. (5). If we also assume that both $n$ and $n'$ are in the (interesting) range of 3-10 (or the corresponding negative range), then the first term on the left-hand-side of Eq. (5) is usually smaller than the second. (Exceptions occur in a small parameter range.) Moreover, the constant term on the right-hand-side of the equation is of order unity. With these simplifications, the equation can be written as,

\begin{equation}
-\left( \frac{1}{n} - 1\right) (\phi_{r1})^2 \sin{(\phi_{r1})} 
\approx \left( \frac{q}{\gamma} \right)^{1/n^{''}}.
\end{equation}

\noindent
With the adopted constraints on $n$ and $\phi_{r1}$, the left-hand-side is always positive and greater than about 2.6 (e.g., when $n = 3$ and $\phi_{r1} = 3\pi/4$). If $n''$ is negative, then this equation can only be satisfied when $q \ll \gamma$. If $\gamma$ is viewed as a core radius (see below), then there can only be a large ring ratio at very small radii in this limit. If $n''$ is positive, then the equation can be satisfied at  $q > \gamma$. However, with the adopted constraints we expect that $1/n'' < 1$, and usually much less than 1.0. Thus, we usually expect that this equation is satisfied, and we can get relatively large ring ratios, only when $q \gg 2.6\gamma A_{\gamma}$ (with the constant from Eq. (5) restored for a better estimate). Since galaxy discs usually don't extend much beyond a few core radii, this limit is also unphysical unless $A_{\gamma}$ is very small. Thus, regardless of the sign of $n''$, the ring ratio is unlikely to exceed 3.0 by very much over the interesting range of radii in the primary disc. 

An optimal exception to this generalization would have $n = 3$ (steeply rising primary rotation curve), $n' = -3/2$ (centrally concentrated companion), and $n'' = 1$. However, these values push our limits and we emphasize again that such exceptions occupy a small area of the relevant parameter space. In most realistic cases the ring ratios will be similar to the FRC case.

This discussion covers the question of maximal ring size ratios; minimal ring size ratios require that the ring middle (RCC) phase be greater than the relevant odd multiple of $\pi$ for the given ring number. By an argument analogous to the one above for the lowest phase, I conclude that the largest physically meaningful phase for the centre of the first ring wave is $3\pi/2$. Then the lowest value of the first-to-second ring size ratio (assuming $\phi_{r2} = 2\pi + \phi_{r1}$) is about $(7/3)^{2/3} \approx 1.8$ (with the same restrictions on the values of $n$ and $n'$ as above). Again, it would be difficult to push to this limit, except in special cases. 

\subsection{Approximating ring sizes in specific systems}

The discussion of the previous subsections suggests a sequence of successive approximations for using Eq. (5) to estimate ring spacings. The first approximation is simply to use Eqs (8) and (9), which are based on the assumption that all terms of Eq. (5) are very small. This will be a good approximation in the case that both collision partners have rotation curves that are close to flat. Unfortunately, the rotation curves of very few ring galaxies, and as far as I know none of their companions, have been measured. If the rotation curves are not very flat, then this can be a rather bad approximation. 

A second approximation is like that used in Eq. (11) where we assume that the first term on the left-hand-side of Eq. (5) is negligible compared to the second. This is valid if, for example, $n'$ is large (positive or negative). We do not, however, want to assume as in Eq. (11) that the coefficient of the right-hand-side of Eq. (5) is of order unity. (E.g., as it is not for large $n$ and $n'$.) Rather we can look at the ratio of the versions of Eq. (5) for two successive rings. In that case the right-hand-side reduces to a power of the ratio of unperturbed ring radii, e.g., $q_{r1}/q_{r2}$. Using a measured ring size ratio, Eq. (10) for $\phi_{r2}$ this equation can be solved for $\phi_{r1}$ (assuming we know the values of $n$ and $n'$, or that they are very large).

This approximation also assumes that the measured ring size ratio $r_{r1}/r_{r2} \approx q_{r1}/q_{r2}$. If this is not the case, it could be refined by successive approximations using the derived values of $\phi$. This approximation will be applied to the Arp 10 ring galaxy in the next section. 

A final approximation is not to neglect either of the terms on the left-hand-side of Eq. (5), but to use the ratio of the equation for two rings, as in the second approximation, to simplify the right-hand-side. Again Eq. (10) and an estimate of the ring size ratio $q_{r1}/q_{r2}$ is used, but the resulting equation for $\phi_{r1}$ is much more complex.

In the case of both the second and third approximations, after $\phi_{r1}$ is determined we can then apply Eq. (5) to the first ring and solve for the amplitude $A$. Interestingly, in many cases, there are a large range of solutions that fail at this point because they yield negative values of $A$. In specific systems values of $A$ can also be constrained by kinematic observations. 

In sum, we see that while the caustic ring wave equations of ASM are very well defined when we know the values of the several parameters, the situation is more complex when the observables provide less information. However, we can get some interesting limits in many cases. We will consider the application to specific systems in the next section. 

\subsection{Ring expansion speeds in the FRC case}

We conclude this section with some simple kinematic results. In the text above Eq. (8) we gave an expression for the epicyclic frequency in an FRC disc. We can multiply that expression by time $t$ and apply it to a ring RCC to get the ring radius as a function of time and the central ring phase. For the $i$th ring we have,

\begin{equation}
r = v_{ri}t, \ \ \ 
\mbox{where,}\ 
v_{ri}  = \frac{\sqrt{2} v_{\gamma}}{\phi_{ri}}.
\end{equation}

\noindent
$v_{ri}$ is the ring velocity, defined as the slope of the $r-t$ relation; $v_{\gamma}$ is the scale velocity given by Eq. (4), which we can identify with the circular velocity in the FRC case. The second equation of Eqs. (12), gives a very beautiful result, that the ratio of the ring velocity to the scale (or circular) velocity is a constant divided by the ring epicyclic phase. Specifically, for a first ring with phase $\phi_{r1} = \pi$, the ring expansion speed is about 0.45 times the circular velocity. 

Eqs. (12) also imply that the ratio of successive ring velocities equals the inverse of the ratio of the ring RCC phases. Thus, according to Eq. (10), in the FRC case, the ratio of successive ring velocities equals the ring size ratio, $R_i$.,

\begin{equation}
\frac{v_{ri}}{v_{r(i+1)}} = \frac{q_{ri}}{q_{r(i+1)}} = R_i.
\end{equation}

\noindent
Note that in the FRC case the actual ring radius $r_i$ equals the initial value $q_i$.

These basic predictions of the analytic theory can be checked against observation in any system. Alternately, given the current paucity of kinematic observations they can be used to predict relative ring speeds from high-resolution images. Examples and more details are given in the following section. The generalization to non-FRC cases is discussed in the Appendix. 

\section{Confronting theory and observation}

In the previous section I showed that in the symmetric, perturbation limit, most cases of physical interest have ring size ratios very close to the fixed values of the FRC case. In Sec. 2, we looked at the available data on ring sizes or upper limits. At the present time, this data is limited, and in many cases of minimal quality. Nonetheless, the comparison with the theory is favorable. First of all, there are no ring size ratios greater than 4.5, well below the theoretical maximum of about 5.0. Similarly, there are none below about 1.7. In the case of second and third rings we would expect some lower values, but there are probably few of these in the sample (though they could be mistaken for first and second rings, see Sec. 4.3).

Secondly, for the measured values, as opposed to the limiting values, there appears to be concentrations at values slightly greater than 3.0 and slightly less than 2.0. The latter few values include the inner second and third rings of the Cartwheel, so we know that they are not first rings. At the present time, the statistical significance of these apparent results is hardly worth calculating. Nonetheless, the data offer no contradictions to the theory, and the theory's predictions are sufficiently definite to compare to better data in the future. There are a few individual cases that merit more detailed discussion now.

\subsection{The case of Arp 10}

\citet[henceforth BMV]{bb7} have done a detailed imaging and Fabry-Perot kinematic study of the Arp 10 ring galaxy. The latter analysis is of higher resolution than the earlier 21 cm kinematic study of \citet{bb10}. The outer ring is not as symmetric as we would like for comparison to the theory. The inner second ring is not clearly separated from the core or bulge light in the broad band imagery. The H$\alpha$ image of BMV does not show it as a very distinct ring. From the Mt. Palomar image of the Arp Atlas I estimate that the first and second rings have a size ratio of $> 3.3$. Evidently based on the H$\alpha$ image, BMV favor a value $> 4$. 

Since BMV find that Arp 10 has a flat rotation curve, we would expect values of the ring size ratio to be close to 3.0. The higher values are somewhat surprising. According to the theory they are certainly too high for the outer ring to be a second ring. This is in agreement with the numerical model of BMV. 

Are the ring ratio estimates so high that they contradict the theory? To answer this question we apply the theory in the second approximation discussed in the previous section. To keep this calculation relatively simple, we begin with the assumptions that the ring size ratio is 3.3 (the lower value), and that the rotation curve of the primary is flat ($1/n \approx 0$). Then, the specific form taken by Eq. (10) in this case is,

\begin{equation}
\frac{q_{r1}}{q_{r2}} = 3.3 
= \frac{\phi_{r2}}{\phi_{r1}}.
\end{equation}

\noindent
The ratio of the two versions of Eq. (5) for the first and second rings reduces to,

\begin{equation}
\frac{2 \cos{\phi_{r1}} +n' {\phi_{r1}} \sin{\phi_{r1}}}
{6.6 \cos{(3.3{\phi_{r1}})} + 10.89 n' {\phi_{r1}} \sin{(3.3{\phi_{r1}})}}
= 3.3^{-2/n'}.
\end{equation}

\noindent
In the second approximation we neglect the cosine terms, and this reduces further to,

\begin{equation}
\sin{\phi_{r1}} = 10.89 \sin{(3.3\phi_{r1})}.
\end{equation}

\noindent
Given the relatively large coefficient on the right-hand-side, the sine term on that side must be small, so $3.3\phi_{r1} \approx 3\pi$. Specifically, $\phi_{r1} = 2.848$, so the second approximation yields a result about 10\% smaller than that of the first approximation $\phi_{r1} = \pi$. 

If we use these results in Eq. (5) for the first ring we find that $n'A \approx 0.85$. Thus, if $n'$ is large (e.g., $n' = 10$), then the amplitude $A$ is small (e.g., $A = 0.085$). If we redo the calculation with an assumed ring ratio of 4.0, we find $\phi_{r1} = 2.345$, and $n'A \approx 0.51$, somewhat smaller. Given the small appearance of the companion in this system, and the high relative velocity inferred by BMV, these estimates seem plausible, at least plausible enough to justify a further quantitative estimate. 

Equations (3), (4), and (5) can be combined to yield,

\begin{equation}
A_{\gamma} = \frac{1}{\sqrt{2}} 
\frac{v^2_{cir2}}{v^2_{cir}}
\frac{v_{cir}}{v_{rel}}
= \frac{1}{\sqrt{2}}
\left( \frac{M_2}{M_1} \right)  _{q=\gamma}
\frac{v_{cir}}{v_{rel}}
\end{equation}

\noindent

BMV estimate the mass ratio of the two galaxies at 1/4, the mean circular velocity at a bit less than $300\ km\ s^{-1}$, and the relative velocity at impact of about $800\ km\ s^{-1}$ (including a slowing since impact to the observed $480\ km\ s^{-1}$). Combining these numbers we get an observationally based estimate of  about $A_{\gamma} = 0.066$, which is very consistent with the analytic estimates for a value of $n' \approx 10$. This estimate is also consistent with the observation that the companion is much less luminous than the primary. 

Using the results of Sec. 3.6 the analytic model can also be compared to the observed kinematic values. Using Eqs. (12) and (13) we find that when the ring size ratio is 3.3 ($\phi_{r1} = 2.848, v_{\gamma} = 300 km\ s^{-1}$), the outer and inner ring expansion velocities are (150, 45) $km\ s^{-1}$. When the ring ratio is 4.0 ($\phi_{r1} = 2.345$), the two expansion velocities are (180, 45) $km\ s^{-1}$. BMV find that the radial velocities of emission line regions in the outer ring range from 30-110 $km\ s^{-1}$, and those in the inner ring are about $25\ km\ s^{-1}$, both significantly lower. 

However, there are several reasons to suspect that the comparison is not so direct. Firstly, there is some uncertainty about the inclination of the Arp 10 disc, see BMV and \citet{bb10}. Moreover, little is known about the warping of that disc, which could help account for the range of velocities seen in the outer ring. Another important factor is that the measured emission regions are very likely to have a different epicyclic phase, and thus a different radial velocity, than the RCC. They could either just be in a modestly different phase of their epicyclic motion, or dissipative interactions could have changed that phase and their radial velocity. Shocks are likely for gas clouds in the ring. \citet{bb29} make similar arguments for the idea that emission line observations yield underestimates of ring speeds, based on their observational and numerical studies of the Cartwheel.

Interestingly, BMV provide another way to estimate the ring speeds. This is based on the strength of a number of specific spectral lines, which depends on the admixture of stellar populations at each radius. The age of the younger populations, whose formation was induced by the rings, depends on the time since ring passage. Given this age and positional information, BMV were able to derive a best-fit kinematic model to the spectral index data. The ring velocities in this model were (180, 46) $km\ s^{-1}$. These velocities have substantial uncertainties (see BMV), however, their agreement with the analytic results is excellent. 

Granted the considerable uncertainties, the simple analytic model provides a good match for the Arp 10 observations. Nonetheless, more accurate data would be very helpful in firming up our understanding of the dynamics of the Arp 10 interaction.

\subsection{The unique, mysterious, and atypical Cartwheel}

The Cartwheel is a more complex case for several reasons. First of all, the size ratio between the first and second ring is the largest known among classical ring galaxies. This is one of the features that makes it so visually spectacular, but also severely constrains the analytic models. Secondly, $21 cm$ kinematic observations \citep{bb16} indicate a rising, not flat, rotation curve. This means the analytic models are more complicated, as well as further constrained by this fact. Thirdly, based on the radial velocity profile given by Higdon, the perturbation amplitude is evidently quite strong. This means that the azimuthal velocities are also significantly affected by the collisional perturbation, so the rotation curve may not give an accurate impression of the gravitational potential. E.g., the ÔtrueÕ rotation curve may be somewhat flatter. This case pushes the limits of the perturbation approximations that the analytic models are based on. That said, it is still worth trying to fit a model, both to see how the analytic models perform at their limits, and to see the differences between this case and FRC cases. 

The second and third items of the previous paragraph suggest opposite approaches. In accord with the second item, I initially explored rapidly rising rotation curve models for the Cartwheel disc. However, it proved essentially impossible to produce a satisfactory model. I then explored models with more moderate rotation curves. Though possible these models are still quite strongly constrained. 

Before continuing we should note that \citet{bb1} presented a Fabry-Perot study of the H$\alpha$ emission in the Cartwheel and companion G2. Their results differ in several ways from Higdon's. Most significantly, while their rotation curve for the approaching side of the Cartwheel disc is quite similar to Higdon's, that for the receding side is much flatter. They observed relatively little emission in the region between the rings, so their rotation curves are much more uncertain in that critical region. They also adopted a mean inclination that is $10^{\circ}$ different than Higdon's. Both papers consider different possible potential centre points, another source of uncertainty. The primary lessons we take from these results are that there are likely to be local velocity variations within the disc beyond those implied by circular rotation, radial ringing, and a modest amount of warping. Given the currently available data, substantial kinematic uncertainties remain. 

As in the case of Arp 10, we start with an estimate of ring size ratio. Here I take the value $R_1 = 4.3$. (\citealt{bb16} obtains a larger value of 4.4, which is within our measurement uncertainties.) A value somewhat larger than that of Table 1 is chosen for reasons that will become apparent at the end of this example. I also set the values $n = n' = 6$. This choice is close to the smallest value of $n$ which yields a value of $\phi_{r2}$ between $(5/2)\pi$ and $3\pi$, when $\phi_{r1}$ is between $\pi/2$  and $\pi$, and the ring ratio is as large as assumed. These conditions are also required for a solution to the equations below. The value of $n'$ is less constrained, so the present choice is somewhat arbitrary.

We will use the third approximation in this example, because the first and second are not sufficiently accurate. In this case, the RCC equation (i.e., Eq. (5)), becomes,

\begin{equation}
\frac{\phi_{r1}}{3} \cos{\phi_{r1}} + 
\frac{5\phi_{r1}^2}{6}  \sin{\phi_{r1}}
= \frac{1}{5A_{\gamma}} 
\left( \frac{q_{ri}}{\gamma} \right)^{-1/6}.
\end{equation}

\noindent
Using Eq. (10) for the phase of the second ring, the ratio of RCC (ring middle) equations for the first and second rings (corresponding to Eq. (15) in the Arp 10 case) is,

\begin{eqnarray}
\nonumber
\frac{ \cos{\phi_{r1}} + 2.5{\phi_{r1}} \sin{\phi_{r1}}}
{5.7565 \cos{(5.7565{\phi_{r1}})} + 82.84 {\phi_{r1}} 
\sin{(5.7565{\phi_{r1}})}}\\
\nonumber\\
= 4.3^{-1/6} = 0.784,
\end{eqnarray}

\noindent
where a very precise value of the ratio of phases (5.7565) is used because of the sensitivity to the sine term in the denominator. The solution of this equation is $\phi_{r1} = 1.623$, which can then be substituted into Eq. (18). However, we cannot immediately solve this for the amplitude, because there is also a dependence on $q_{r1}$. Specifically, Eq. (18) reduces to

\begin{equation}
\frac{q_{r1}}{\gamma} 
= \left( \frac{0.0497}{A_{\gamma}} \right) ^6.
\end{equation}

\noindent
The length $\gamma$ is a free parameter from the mathematical point of view. Physically, it was used by ASM for two purposes. First, it was introduced as a scale length in the rotation curve of the ring galaxy. In this context it is natural to assume that it is the smallest radius at which the power-law rotation curve applies. One expects deviations in the galaxy core, so $\gamma$ can be viewed as a core radius. Secondly, $\gamma$ was also used as a scale length in the radial dependence of the perturbation amplitude. Use of the same scale length was a simplification. As can be seen from Eq. (2) and the second of Eqs. (5), the radius dependent perturbation depends on the structure of both the companion and the primary. The double use of the same scale length is justified only if both galaxies have about the same scale length, e.g., core size. We will continue to make that assumption for convenience. 

As discussed above, if $\gamma$ is a core radius, we generally expect that the ring radii will be of the order of several core radii. Another aspect of the identification of $\gamma$ with the core scale is that it implies that the radius of the extremum of the perturbation is at about the same radius, which seems at least physically plausible. The point here is that an analytic model that requires, say ${q_{r1}}/\gamma \gg 10$ or ${q_{r1}}/\gamma \ll 0.1$ for reasonable amplitudes $A_{\gamma}$,  is somewhat unphysical.

To meet that requirement, we see that Eq. (20) suggests a value for $A_{\gamma}$, of slightly less than 0.0497. Specifically, for $A_{\gamma}$ in the range 0.035 - 0.040, ${q_{r1}}/\gamma$ is in the range 3.7 - 8.2. The latter range seems reasonable, so we adopt the former range for the amplitude. 

Because we clearly observe the second ring in the Cartwheel it is worth computing a few more things. First recall that the ring radii are not actually the $q_{ri}$, but rather $r_i = q_{ri}(1 - A(q) \sin{(\phi_{ri})})$. Using the values derived above we find that $r_1/r_2  = 4.2$, little different from the assumed value of $R_1$, but in agreement with Table 1. 

The HST imaging of the Cartwheel reveals ring/spiral structure inside of the second ring. In Table 1 I have identified this as a third ring and estimated its size, despite its uncertain nature. The calculation above can be repeated to derive the parameters of a third ring in this sample model. The specific form of the RCC equation and other details are straightforward extensions of those above, except that in this case we wish to solve for the ring ratio ($R_2$) rather than the phase of the outermost of the pair (which is now the given quantity). We find the $\phi_{r3} = 4.981\pi$ and $R_2 \approx 1.5$. Table 1 shows that our estimate from the image is about 1.8. Given that the third ring is not well formed, that we know nothing about the rotation curve in the innermost regions, and that the collision was probably not symmetric on these very small scales (about 2 kpc) the discrepancy between these two numbers is not great. The result lends some support to the notion that the inner arc is indeed part of a third ring.

At this point our example model seems relatively successful in comparison to the observed ring morphology, but we should compare its predicted perturbation amplitude to the observations. \citet{bb16} estimates that the mass of each of the individual companions is less than 6\% that of the Cartwheel, which sounds quite close to our amplitude estimate. However, the comparison is not that simple. Higdon's indicative masses are derived from formula of \citet{bb6}, which uses HI line widths and an optical disc size. (A kind of early Tully-Fisher relationship seems to be the basis of this formula.) If the distant companion G3 is the ring-making collision partner, as suggested by Higdon, then we note that its HI distribution does not extend much beyond the fairly compact optical disc, and Higdon believes that most of its gas (and stars?) has been removed to form a long bridge to the Cartwheel. In that case, it seems quite possible that as a result of the encounter both factors in the mass formula may underestimate the mass of the companion halo, which perturbed the Cartwheel disc. Companion G2 does have an extensive HI disc, so the indicative mass may be more reliable if it was the collision partner. However, its mass is less than half that of G3, and would seem to be too low to be the sole cause of the Cartwheel's rings. Companions G1 and G2 may be interacting with each other, and may have both collided with the Cartwheel. Their combined indicative mass is close to that of G3. G1 has no detected HI gas, and its mass estimate is based on its optical velocity dispersion, which is likely to underestimate the mass of an extended halo. 

In sum, it seems possible that the mass ratio of the collision partners was up to a few times larger than the 6\% estimate of \citet{bb16}. According to Eq. (17), the perturbation amplitude also depends on a numerical factor and the ratio of the Cartwheel's circular velocity to the collision velocity. Higdon estimates the current relative velocity of G3 and the Cartwheel to be about 360 $km\ s^{-1}$. As in the case of Arp 10 it was probably considerably higher at the time of impact. The Cartwheel's circular velocity at the ring is close to 300 $km\ s^{-1}$ (\citealt{bb16}, Amram et al.Õs value is about 10\% lower), which implies a velocity ratio in Eq. (17) of about 0.5. The factor is roughly the same if G1 and G2 were the collision partners. In the case of our example model, the numerical factor in Eq. (11) is about 0.5,  instead of $2^{-1/2}$ as in the FRC case (see below and Appendix). Combining factors gives an observational amplitude estimate of 6\% times a few times two factors of 0.5, which is very close to the model amplitude $A_\gamma$ of about 4\%. The only problem is that in using the total masses and the circular velocity at the ring radius I have effectively computed the amplitude at the ring. The model amplitude at that radius is about 0.04 x 3.7 = 0.15 (or alternately $0.045 \times 1.8 = 0.08$). Given the large uncertainties, these numbers may be consistent, but at this point the comparison is not reassuring.

As in the case of Arp 10 we can look to the kinematics of the Cartwheel for more information. For the present model, the ring speed equation, analogous to the second of Eqs. (12) is, 

\begin{equation}
 \frac{v_{ri}}{v_{\gamma}}
= \frac{1.83}{\phi_{ri}}
\left( \frac{q}{\gamma}  \right) ^{1/6}.
\end{equation}

According to \citet{bb16} the outer ring radius is at about $16.5\ kpc$. With the factor $q_{r1}/\gamma \approx 3.7$ from above we have $\gamma \approx 4.4\ kpc$ (only slightly larger than the radius of the second ring). At that radius, Higdon's rotation curve gives us a circular velocity of about $90\ km\ s^{-1}$ (or up to $190\ km\ s^{-1}$ in \citealt{bb1}). Using these values, and the value for the phase derived above (1.623), we get an expansion velocity of $126\ km\ s^{-1}$ for the outer ring (or up to $170\ km\ s^{-1}$ for Amram's receding side). Higdon estimates a value of about $50\ km\ s^{-1}$. Noting the tendency for ring speeds derived from emission line velocities to be low, \citet{bb29} suggest a value of close to $100\ km\ s^{-1}$. In any case our model value is too high, though not by much in the latter case. The real problem with the kinematics is that it is in stark contrast to HigdonÕs HI rotation curve. Given the value of the circular velocity at the $r = \gamma$, the model predicts a value of $110\ km\ s^{-1}$ at the ring radius. Higdon finds a value close to $270\ km\ s^{-1}$. His rotation curve is fairly steeply rising at all radii, while the model is much flatter, and the formalism cannot account for such a large ring ratio with such a rising rotation curve. Amram et al.Õs receding side rotation curve is more consistent with the model, except for a large predicted ring speed.

On balance the model is not too bad. It is consistent with the large ring spacing, and the development of a third ring, like the observed inner arc. However, it suggests a perturbation amplitude that is too large at the outer ring radius, and it fails to match the observed kinematics. Steeply rising rotation curves in all or part of the disc are not common in galaxies, though low surface brightness galaxies often have moderately rising ones. Such rotation curves are more common among disturbed galaxies, e.g., see the study of Virgo spirals by \citet{bb23}.

This kinematic failure may provide a clue to the Cartwheel's development. The models would be less constrained and the amplitude discrepancy could be relieved if the rotation curve was flatter throughout and the first two rings were not so widely spread. I conjecture that this was the case when G3 hit the Cartwheel and started the formation of the rings, but that a second, prograde and off-centre encounter with G1/G2 disturbed the kinematics in a manner like that of some of the Virgo spirals. Parts of the outer disc may have been spun up, and outward, helping to account for the large diameter of the outer ring. 

There are several collateral facts in support of this picture. First the HI bridge to G3 is long, suggesting that it has been some time since impact (\citealt{bb16}). It traces back to nearly the centre of the Cartwheel, and to produce three nearly symmetric rings the impact must have been very near the centre. Secondly, the X-ray bridge towards G1/G2 is much shorter (\citealt{bb14}), suggesting a more recent encounter. The extensive gas disc of G2 would likely have been destroyed in a direct encounter with the Cartwheel disc, so it probably did not collide directly with that disc. The mass of G1 seems too low for it to have produced the rings alone (though it may be underestimated). It seems much more likely that this pair hit the outer disc of the Cartwheel (each galaxy at somewhat different times), possibly removing gas from G1. The Cartwheel's rings are elliptical, but their major axes point in different directions, and so the ellipse orientations are not simply the result of projection onto the sky. They may be the result of disc warping, but that is less likely for the two innermost rings. Radius dependent torques in a second encounter may be a better explanation. Finally, we note that the unusual spokes would be much easier to account for in a scenario with external torques. 

It will take detailed numerical modeling to test this two-hit conjecture. While it is more complicated than a single-hit picture, it does seem an economical hypothesis for accounting for a range of Cartwheel peculiarities. Most obviously it would account for the rarity of such a large ring spacing. Another comparable case in Table 1 is NGC 2793, one of the most asymmetric rings of the collection, so perhaps also a victim of strong torques. KUG 445B, object 8 in Table 2, has a large ring spacing and is not asymmetric. However, it has a significantly larger neighbor. If this is the collision partner, then the disturbance may have been very nonlinear ($A_{\gamma} >> 1$).

A number of numerical models of the Cartwheel have been published in the last twenty years. None of them include two collisions, nor account for the range of phenomena discussed above. (In fairness, most of them were designed only to fit a subset of Cartwheel phenomenology, e.g., spokes or multiple rings.) Specifically, the models of \citet{bb15}, \citet{bb9}, and \citet{bb17} do not reproduce the ring spacing. Those of \citet{bb27} do, but have a steeply rising rotation curve in a rigid halo potential. Their outer ring is very weak, which may be another reflection of the difficulties in making an analytic model with a steeply rising rotation curve. \citet{bb29} present a two-dimensional model with similar characteristics. None of these models attempts to match the shape of the Cartwheel rings. The structural complexities of the Cartwheel, and the dynamic complications suggested by the scenario above, are likely to continue to challenge numerical modelers for some time.

In this case we have learned a couple of things about the analytic models. First, although they have a number of adjustable parameters, to paraphrase Pauli, they are good enough to be proven wrong, or at least inadequate. Secondly, the way a model fails, or even the difficulties in getting a fit, can provide clues to what is missing from the model. This information can be very complementary to that provided by numerical models.

\subsection{AM2136-492 (ESO236-29) and other closely spaced rings}

As a third example consider the AM2136-492 system, which has two large rings that are much closer together than the previous two examples. This system has the smallest measured size ratio in Table 1 (1.8), and it is nearly impossible to see the rings as the first and second ones within the context of the assumptions underlying the analytic theory. Of course, they may not both be the result of a collision. However, there is no bar component to hold responsible for the inner ring. Nonetheless, this could be a resonance ring which has persisted after the dissolution of a bar. On the other hand, the enormous physical sizes of these rings argues against this (see Table 1). MNP identify a possible companion at about two diameters projected separation. This is about the separation we would expect if the two rings were in fact the second and third rings. The analytic models and comparisons to other systems in Table 1 both suggest that this possibility is much more consistent with the small ring spacing, assuming the rings are the result of a collision.

This hypothesis immediately raises a few questions. Firstly, where is the first ring? If it were typically more than three times the size of the outer ring, it would extend most of the distance to the putative companion, and thus, may well have propagated off the primary disc. If not, it may be too faint to detect on the survey image. The present outer (e.g., second) ring is already quite faint. This raises the couple questions Ð why is the second ring so faint and why is the third ring so thin? In typical analytic models of ASM the opposite of both circumstances is usually true. In the MNP sample faint outer rings are in the minority. 

At the least it would seem to require a very weak perturbation, and thus, a small or fast companion. In the present case, the companion is not so distant that we have any reason to believe that the relative velocity is especially large. On the other hand, the companion appears significantly less luminous than the primary. If it is experiencing an enhanced rate of SF (unknown at present), then the mass ratio may be even smaller than the luminosity ratio. In sum, the observations are limited, and quantitative tests would require kinematic data.

The apparently strong SF in the third ring, together with its thinness, raise another issue. It is likely that the third ring is the first caustic ring, or nearly so, with actual stellar orbital crossings in radius, rather than just orbit compressions. The gas compressions associated with caustic formation, which is qualitatively like the breaking of a wave, are much stronger than those associated with mere orbital compressions. This could account for the differences in SF in the two rings. 

Strong caustics are associated with wide stellar rings, though the gas may still be found in thin shocks (see ASM). We see very few noticeably wide stellar rings in the objects of Table 1. For outer rings this may mean that they have propagated beyond the stellar disc in many cases. For inner rings, width compromises our ability to see them as distinct rings until they propagate out of the core, so there may be a number of indiscernible ones. In any case, it appears that many of the outer rings in the sample are either not caustics, or just barely satisfy the orbit-crossing condition. Given that, one may wonder about the validity of the theory, since the relevant equations are ``caustic'' equations. Based on the success of the theory in providing reasonable models for real systems it appears that the caustic RCC equation does provide a valid representation of the compression centre, even when the wave does not ``break.'' 

Another possible member of an AM2136-492 class is the $z = 0.5$, radio galaxy 16V31 studied by \citet{bb21a}. The faintness and low resolution of the HST WFPC2 image of this distant object make it hard to be completely confident that it is a colliding ring galaxy. The distance and lack of a visible candidate companion would disqualify it from inclusion in the MNP catalog. However, it may have two symmetric and similar rings, and with the sizes given by the above authors, the size ratio is 1.8, like AM2136-492. In contrast to the latter, in the rings have similar surface brightnesses, with the outer one being somewhat brighter. The possibility that these are again second and third rings is consistent with the companion having had time to move far enough away to be off the image. To learn anything more about this system would require the sensitivity and resolution of HST ACS or WFPC3 imaging.

Another similar system is ESO 381-47 recently studied with radio, optical and UV observations by \citet{bb12}. In their disc and bulge subtracted residual image there appear to be three nearly circular rings. The outer ring is star-forming in the outer gas disc of the galaxy. The middle ring is apparently the widest, though the inner ring may not be fully formed, and may be comparably wide. The size ratios of both the outer-to-middle and the middle-to-inner are about 2. This is interesting, because it would imply that the outer ring is the second ring and the inner ring is the fourth. The galaxy is part of a group, with several possible companions. They all appear to be considerably less massive, and so, could be responsible for a weak perturbation, which gave rise to a system of weak, thin rings. Because the rings appear to be nearly face-on there is little kinematic information, including evidence for outward propagation of the ring waves. Donovan et al. consider a number of possible causes for the rings besides a collision, including accretion of the outer disc gas. Given this considered uncertainty, I have chosen not to include it in Table 1. 

And finally, another possible system is NGC 1961, as described by a recent preprint by \citet{bb11}. The visible ring ratio is about 2. It appears possible that another inner ring is forming. The collisional model provided by Combes et al. suggests that we are looking at a first and second (and possibly third) rings. The galaxy is quite asymmetric, so the analytic theory may not be applicable. 

\subsection{Mapelli et al.Õs rings}

Recently, \citet{bb18a} have shown that the properties of some well-known giant low surface brightness (GLSB) galaxies can be accounted for as the result of what we might call extreme ring-making collisions. What I mean by this term is that in their simulations the companion is massive, more than half the mass of the primary, and significantly more compact, so the effective mass ratio contribution to the perturbation amplitude in the disc is probably more than unity. The relative velocity at impact is the escape velocity and the rotation curves are roughly flat, so Eq. (17) applies, and we can estimate the amplitude as $A \ge 0.5$. 

I have not included Mapelli et al.Õs GLSB galaxies in Table 1 because the objects are very unusual compared to other ring galaxies. However, Mapelli et al.Õs comparison between observations and numerical models is convincing. The physical size of these rings is very large compared to resonance rings, for example. Thus, the interpretation that these galaxies are rings seems likely, even though in the models their structure looks unusual compared to other ring galaxy simulations. On close inspection, it appears that the outer ring in these models is very broad, in fact, extending across more than half the disc. This, and a huge expansion of the disc as the ring wave reaches its outer parts, is what the analytic theory would predict for a large amplitude perturbation. This situation is in stark contrast to that of the thin rings of ESO 381-47 discussed in the last subsection. 

The analytic theory is not really valid in the limit of large amplitudes. E.g., the epicyclic approximation to the orbits is probably not accurate. However, we can again test the limits of the theory and its usefulness by applying it to a Mapelli-type example. To begin, consider the caustic equation, i.e., the equation that determines the inner and outer caustic edges. It is (see ASM),

\begin{equation}
\left( 1 - \frac{1}{n''} \right)  \sin{\phi} +
\left( \frac{1}{n} - 1 \right) {\phi} \cos{\phi} 
= \frac{1}{A_{\gamma}} 
\left( \frac{q}{\gamma} \right)^{1/n''}.
\end{equation}

\noindent
In the present case we assume the $n$, $n'$, and $n''$ all are large, and that $A_{\gamma} = 0.5$. Then the equation becomes simply $\sin{(\phi)} - \phi \cos{(\phi)} = 2$. The solutions for the first ring are $\phi = 2.15$ and 3.95. With these epicyclic phases we find that the inner caustic edge is $r_{min} = 0.58q$, the RCC is about $r = q$, and the outer edge is $r_{max} = 1.36q$. Thus, the ring width $\Delta r/r = 0.78$ in this example. Certainly a ring that is most of the disc when it propagates to the outer parts merits the description Òextreme.Ó Nonetheless, the analytic theory still provides a useful approximation.

With such a large perturbation we would not expect these rings to be planar, and the numerical models confirm a large warping. This means that gas clouds can pursue their epicyclic oscillations without colliding in the ring with their counterparts at very different initial radii, just like the stars. Local, low velocity collisions, between clouds are still possible. Thus, we do not expect the gas to pile up at a strong shock front, and star formation to occur throughout the broad ring. This is what Mapelli et al.Õs simulations show, albeit with some pileup at the caustics (see their Fig. 8). These galaxies are well worth further detailed study.

Finally, I would note that the analysis above could be done in reverse for an observed system, like that of previous systems. That is, with observed values of $r_{min}$, $r_{max}$ and the ring centre ($q$), the edge phases and the amplitude could be estimated from specific instances of Eq. (22) and the epicyclic orbit equation (for the ring centre radius). The derived amplitude could be compared to the observed companion as a model check. With observationally derived kinematic information, the indices $n$ and $n'$ would not have to be assumed. With observations of a second, inner ring and the ring spacing, precise ring centre phases could be derived, as well as obtaining redundant information on the amplitude to check the model consistency. All of which is to say that information on the positions, widths and kinematics of multiple rings would completely specify the analytic model for any system, and provide redundant checks. 

\subsection{Andromeda shows how to break the rules}

\citet{bb8} have recently proposed that M31, the Andromeda Galaxy, might be a colliding ring galaxy, with both an inner and outer ring easily visible in Spitzer infrared observations. Because of its proximity this should be an ideal object for detailed study. Unfortunately, the numerical model presented by Block et al. indicates that it is not. The Andromeda rings have a size ratio of about 7, well outside the range predicted by the models above for a first and second ring. The animation of the numerical model (provided to the author by F. Combes), shows that in the inner disc the ring waves interact with pre-existing spirals and a bar. The interaction is so strong it makes it difficult to distinguish the second and higher ring waves. 

It is possible that the observed inner ring is a third wave, while the second wave is not visible due to the wave interference. In that case, the large size ratio is compatible with analytic models for the ratio of a first to third ring. However, in the animation it appears that the second ring is almost captured within the inner bar, so it may not be accurate to describe the interaction as simply a disappearing second ring. 

The lesson of this simulation is that in the case of a small companion, generating weak disturbances in a disc dominated by strong pre-existing waves, the ring waves can interact too strongly to propagate in the classical manner. They may not even be visible. The problem is likely exacerbated in the case of off-centre collisions. In survey observations it would be hard to resolve an inner ring with such a small relative size. This may be one reason why other cases have not been documented. Additionally, the outer ring will generally be weak in such cases, making them hard to identify as ring galaxies. Nonetheless, it is a very interesting class of object. 

\section{Summary and conclusions}

This study began in Sec. 2 with the assembly of a sample of a few dozen symmetric ring galaxies, largely drawn from the sample of MNP. Although the amount and quality of data available for most of these objects is very limited, the optical imagery could be used to estimate the size ratio of successive rings, or give limits thereof. The result was a quite restricted range of variation of that quantity and evidence for a bimodal distribution, albeit with few data points and many limiting values (see Fig. 2). Although these results are very tentative, they are in qualitative accord with the analytic models of ring galaxies developed by \citet{bb28} and ASM, and together with detailed observations of a few specific systems, motivated renewed consideration of those models. 

In Sec. 3 the analytic theory was reviewed and specific expressions were presented for ring spacings and ring kinematics. The importance of the FRC case was emphasized, both because it is common, and because the analytic model is particularly simple in that case, and in similar cases with modestly rising or falling rotation curves. The peaks of the observed bimodal distribution correspond to the size ratios of first to second, and second to third rings in FRC discs, according to the analytic theory. 

In Sec. 4 the theory was applied to the study of a few systems which have been extensively observed, and which have multiple rings. The general procedure begins by using the observed ring size ratio and Eq. (10) to derive the ratio of the epicyclic phases of the RCCs. The ratio of the two RCC equations (two versions of Eq. (5)) provides a second equation from which to derive those two phases. The magnitude of the perturbation amplitude cancels in this latter equation. After the phases have been obtained, either of the individual ring RCC equations can be used to derive the perturbation amplitude. This can be compared to fundamental constraints (it should lie between 0 and 1), and to the observed luminosity ratio to judge accuracy of the model. 

In this procedure the form of the rotation curves of the two galaxies is input at the outset, and the accuracy of that initial estimate is judged by the computational outcomes. Most ring size ratios will only allow reasonable models for limited ranges of the rotation curve indexes $n$ and $n'$ and limited ranges of the amplitude $A_{\gamma}$. This point is well illustrated by the examples of Sec. 4. The models are much more constrained when kinematic observations are available, when a third ring is visible, or when caustic widths can be measured. An example of the latter case is provided by the extreme rings of Mapelli et al. (2008).

The Arp 10 system, the first example of Sec 4, could be fit quite well by an analytic model, though there are some complications in the observed kinematics. All of the various observations of the Cartwheel system, the second example, could not be fit very well by an analytic model. The primary difficulties stem from the disc kinematics, and together with other clues, they inspire a two-collision conjecture, described above. The third example, AM2136-492 could also be well fit by an analytic model, but only if the observed rings are the second and third. As discussed in Sec. 4.3 that result tells us some very interesting things about that relatively unstudied system, and caustic rings generally. 

The range of wave strengths from AM2136-492 to MapelliÕs et al.Õs rings is probably very close to the maximum. Perturbations much less than the former case would hardly generate a visible ring. Perturbations much more than the latter case, would be greater than unity and would likely destroy the disc rather than ÔringÕ it. It is encouraging that the analytic models are evidently useful over the whole of this range.

I conclude by emphasizing two points. First, no other type of colliding galaxy, nor system of waves within a galaxy disc can be modeled as simply, with a few algebraic equations, as the systems described in the preceding sections. The symmetry of ring galaxies is special and allows them to be used for fundamental tests of our understanding of the dynamics of galaxy discs.

Secondly, as is evident from the MNP paper and the literature in general, ring galaxies are under-observed. If the data available for all of the galaxies in Table 1 was comparable to that of Arp 10 and the Cartwheel, many fundamental questions could be addressed using the special properties of rings, including the following. What are the typical relative velocities in direct collisions, how many lead to merger, and on what timescale? How do the answers to these questions compare to cosmological simulations? What range of companion masses and collision parameters leads to significant disturbances in the target disc? How does the SF in a ring wave depend on the wave amplitude? How much does it depend on other parameters, like gas surface densities? What kind of star clusters are formed as a function of wave amplitude, and how do they evolve behind the wave? Given the answer to these questions, what is the net SF induced by ring waves in their passage through a galaxy disc? 

A number of these questions are of general relevance to the problem of how galaxy discs evolve and are evolved by waves. In other situations where waves are induced by less symmetric interactions or bars, numerical models can be compared to observations to shed light on these questions. However, for ring galaxies the analytic models offer additional information and checks on the interpretation of the observation. The potential for substantial scientific return seems high for observations of ring galaxy structure and kinematics that are able to resolve multiple rings. 

\section*{Acknowledgments}

I am grateful to B. J. Smith and F. Combes for comments on an early version of this paper, and to Phil Appleton for many years of insights on ring galaxies. I am also grateful to an anonymous referee whose comments contribued considerably. Many thanks also to the Galaxy Zoo(www.galaxyzoo.org) organizers and participants, especially the colliding rings forum group, for their work in finding more objects. This research has made use of the NASA/IPAC Extragalactic Database (NED), which is operated by the Jet Propulsion Laboratory, California Institute of Technology, under contract with the National Aeronautics and Space Administration.

\appendix

\section[]{General ring expansion speeds }

In Sec. 3.6 we considered the ring expansion speeds in the FRC case. The expression for ring speeds in discs with general power-law rotation curves is somewhat more complicated. The discussion of the Cartwheel above provided one example of its use, so I present the general form here for reference.

We can begin by substituting the equation for the epicyclic frequency into the epicyclic phase expression,

\begin{equation}
\phi_{ri} = \kappa t =
\sqrt{2} \left( 1 + \frac{1}{n} \right)^{1/2}
\frac{v_{\gamma} t}{\gamma}
\left( \frac{q_i}{\gamma} \right)^{\frac{1}{n} -1}.
\end{equation}

\noindent
Then we differentiate with respect to time (after slight rearrangement to make sure the explicit time dependence is eliminated), to get,

\begin{equation}
 \frac{v_{qi}}{v_{\gamma}}
= \frac{1}{v_{\gamma}}  \frac{dq}{dt} = 
\frac{\sqrt{2 \left( 1 + \frac{1}{n} \right)}}
{\left( 1 - \frac{1}{n} \right) \phi_{ri}}
\left( \frac{q_i}{\gamma} \right)^{1/n}.
\end{equation}

\noindent
This is the speed with which the wave moves through disc annuli specified by their unperturbed radius $q$. 

Analogous to Eq. (13), an expression for the ratio of the speeds of two successive ring waves is

\begin{equation}
\frac{v_{qi}}{v_{q(i+1)}} = \left( \frac{\phi_{i+1}}{\phi_i} \right)
{\frac{q_{ri}}{q_{r(i+1)}}}^{1/n}
= \left( {\frac{q_{ri}}{q_{r(i+1)}}} \right) ^{\frac{n^2 +n - 1}{n(n-1)}},
\end{equation}

\noindent
where Eq. (10) was used to obtain the last equality. Note that for large $n$ the last exponent is approximately 1, as in Eq. (13), so that simple scaling is approximately retained for nearly flat rotation curves for $q$-radii and $q$-velocities. To obtain the value of the physical radii ($r_i$) and velocities requires solving for $\phi_i (q)$ for the given potential as described in the text. Expressions for these quantities also include amplitude dependences, which introduce further complications. Given the state of the observations, deriving explicit expressions in potentials that differ substantially from the FRC case does not seem justified at present.

\bsp

\label{lastpage}


\begin{thebibliography}{99}

\bibitem[\protect\citeauthoryear{Amram et al.}{1998}]{bb1}Amram, P., Mendes de Oliveira, C., Boulesteix, J.,  Balkowski, C. 1998, A\&A, 330, 881
\bibitem[\protect\citeauthoryear{Appleton \& Marston}{1997}]{bb2}Appleton, P. N., Marston, A. P. 1997, AJ, 113, 201
\bibitem[\protect\citeauthoryear{Appleton \& Struck-Marcell}{1996}]{bb3}Appleton, P. N., Struck-Marcell, C. 1996, Fund. Cosmic Phys., 16, 111 (ASM)
\bibitem[\protect\citeauthoryear{Arp}{1966}]{bb4}Arp, H. C. 1966, Atlas of Peculiar Galaxies, Caltech, Pasadena
\bibitem[\protect\citeauthoryear{Arp \& Madore}{1987}]{bb5}Arp, H. C., Madore, B. F. 1987, Catalog of Southern Peculiar Galaxies and Associations, Vols. I and II, Cambridge University Press, Cambridge
\bibitem[\protect\citeauthoryear{Athanassoula et al.}{1982}]{bb5a} Athanassoula, E., Bosma, A., Cr\'{e}z\'{e}, M., Schwarz, M. P., 1982, A\&A, 107, 101
\bibitem[\protect\citeauthoryear{Balkowski}{1973}]{bb6} Balkowski, C. 1973, A\&A, 29, 43
\bibitem[\protect\citeauthoryear{Bizyaev, Moiseev, \& Vorobyov}{2007}]{bb7} Bizyaev, D. V., Moiseev, A. V.,  Vorobyov, E. I. 2007, ApJ, 662, 304 (BMV)
\bibitem[\protect\citeauthoryear{Block et al.}{2006}]{bb8} Block, D. L., Bournaud, F., Combes, F., Groess, R., Barmby, P., Ashby, M. L. N., Fazio, G. G., Pahre, M. A., Willner, S. P. 2006, Nature, 443, 832
\bibitem[\protect\citeauthoryear{Bosma}{2001}]{bb9} Bosma, A. 2001, in M. Valtonen, C. Flynn, eds.  Small Galaxy Groups, ASP Conf. Series 209, ASP, San Francisco, p. 255
\bibitem[\protect\citeauthoryear{Charmandaris \& Appleton}{1996}]{bb10} Charmandaris, V., Appleton, P. N. 1996, ApJ, 460, 686
\bibitem[\protect\citeauthoryear{Combes et al.}{2009}]{bb11} Combes, F., Baker, A. J., Schinnerer, E., Garc'a-Burillo, S., Hunt, L. K., Boone, F., Eckart, A., Neri, R., Tacconi, L. J. 2009, A\&A, in press (arXiv:0906.2493)
\bibitem[\protect\citeauthoryear{de Vaucouleurs \& Buta}{1980}]{bb11a} de Vaucouleurs, G., Buta, R. J. 1980, AJ, 85, 637
\bibitem[\protect\citeauthoryear{Donovan et al.}{2009}]{bb12} Donovan, J. L., Serra, P., van Gorkom, J. H., Trager, S. C., Oosterloo, T., Hibbard, J. E., Morganti, R. Schiminovich, D., van der Hulst, J. M. 2009, AJ, 137, 5037
\bibitem[\protect\citeauthoryear{Few \& Madore}{1986}]{bb13} Few, J. M. A., Madore, B. F. 1986, MNRAS, 222, 673
\bibitem[\protect\citeauthoryear{Gao et al.}{2003}]{bb14} Gao, Y., Wang, Q. D., Appleton, P. N., Lucas, R. A. 2003, ApJ, 596, L171
\bibitem[\protect\citeauthoryear{Hernquist \& Weil}{1993}]{bb15} Hernquist, L., Weil, M. L. 1993, MNRAS, 261, 804
\bibitem[\protect\citeauthoryear{Higdon}{1996}]{bb16} Higdon, J. L. 1996, ApJ, 467, 241
\bibitem[\protect\citeauthoryear{Horellou \& Combes}{2001}]{bb17} Horellou, C., Combes, F. 2001, ApSpSci, 276, 1141
\bibitem[\protect\citeauthoryear{Lynds \& Toomre}{1976}]{bb18}Lynds, R., Toomre, A. 1976, ApJ, 209, 382
\bibitem[\protect\citeauthoryear{Madore, Nelson, \& Petrillo}{2009}]{bb19} Madore, B. F., Nelson, E., Petrillo, K. 2009, ApJS, 181, 572 (MNP)
\bibitem[\protect\citeauthoryear{Mapelli et al.}{2008}]{bb18a} Mapelli, M., Moore, B., Ripamonti, E., Mayer, L., Colpi, M.,  Giordano, L. 2008, MNRAS, 383, 1223 
\bibitem[\protect\citeauthoryear{Marston \& Appleton}{1995}]{bb20} Marston, A. P., Appleton, P. N. 1995, AJ, 109, 1002
\bibitem[\protect\citeauthoryear{Mart\'{i}nez-Garc\'{i}a, Gonz\'{a}lez-L\'{o}pezlira, \& Bruzual-A}{2009}]{bb21} Mart\'{i}nez-Garc\'{i}a,  E. E., Gonz\'{a}lez-L\'{o}pezlira, R. A., Bruzual-A, G. 2009, ApJ, 694, 512
\bibitem[\protect\citeauthoryear{Roche, Lowenthal, \& Koo}{2002}]{bb21a} Roche, N. D., Lowenthal, J. D., Koo, D. C. 2002, MNRAS, 337, 840
\bibitem[\protect\citeauthoryear{Romano, Mayya, \& Vorobyov}{2008}]{bb22} Romano, R., Mayya, Y. D., Vorobyov, E. I. 2008, AJ, 136, 1259
\bibitem[\protect\citeauthoryear{Rubin, Waterman, \& Kenney}{1999}]{bb23} Rubin, V. C., Waterman, A. H., Kenney, J. D. P. 1999, AJ, 118, 236
\bibitem[\protect\citeauthoryear{Springel et al.}{2005}]{bb24} Springel, V., et al. 2005, Nature, 435, 629
\bibitem[\protect\citeauthoryear{Struck}{2006}]{bb25} Struck, C. 2006, AJ, 131, 1347 
\bibitem[\protect\citeauthoryear{Struck-Marcell}{1990}]{bb26} Struck-Marcell, C. 1990, AJ, 99, 77
\bibitem[\protect\citeauthoryear{Struck-Marcell \& Higdon}{1993}]{bb27} Struck-Marcell, C., Higdon, J. L. 1993, ApJ, 411, 108
\bibitem[\protect\citeauthoryear{Struck-Marcell \& Lotan}{1990}]{bb28} Struck-Marcell, C., Lotan, P. 1990, ApJ, 358, 99
\bibitem[\protect\citeauthoryear{Vorobyov \& Bizyaev}{2003}]{bb29} Vorobyov, E. I., Bizyaev, D. 2003, A\&A, 400, 81
\bibitem[\protect\citeauthoryear{Wallin \& Struck-Marcell}{1994}]{bb30} Wallin, J. F., Struck-Marcell, C. 1994, ApJ, 433, 631

\end{thebibliography}
\end{document}